# The small-scale clustering power spectrum and relativistic decays

S.J. McNally[1] and J.A. Peacock[2]

[1] *Institute for Astronomy, University of Edinburgh, Blackford Hill, Edinburgh EH9 3HJ*
[2] *Royal Observatory, Blackford Hill, Edinburgh EH9 3HJ*

**ABSTRACT**
We present constraints on decaying-particle models in which an enhanced relativistic density allows an $\Omega = 1$ Cold Dark Matter universe to be reconciled with acceptable values for the Hubble constant. Such models may contain extra small-scale power, which can have important consequences for enhanced object formation at high redshifts. Small-scale galaxy clustering and abundances of high-redshift damped Lyman-$\alpha$ absorption clouds give a preferred range for the mass of any such decaying particle of 2 to 30 keV and a lifetime of 0.5 to 100 years for models with a high Hubble constant ($h > 0.75$). A lower Hubble constant, $h \simeq 0.5$, weakens the constraint to $0.5 < m < 30$ keV, $0.2 < \tau < 500$ years. In permitted versions of the model, reionization occurs at redshifts $\sim 10 - 200$, and this feature may be of importance in understanding degree-scale CMB anisotropies.

## 1 INTRODUCTION

Observations of large-scale galaxy clustering have produced something of a crisis for theories of cosmological structure formation. The paradigm for the past decade has been a model where scale-invariant adiabatic primordial fluctuations cause clustering to grow in dark matter which is collisionless, and which has negligible thermal velocities; in many ways, observations have supported the basic elements of this picture. The clustering power spectrum is smooth and featureless, with no sign of the oscillatory features that would be expected if normal baryonic material was dynamically dominant (e.g. Peacock & Dodds 1994; hereafter PD). The detection of microwave-background anisotropies on a variety of angular scales favours a fluctuation spectrum which is indeed close to adiabatic and scale-invariant for wavelengths above about 100 Mpc (e.g. White, Scott & Silk 1994).

Despite these encouraging features, there has emerged a consensus that there is a problem with the shape of the fluctuation spectrum. The density of the universe should be written on the sky in the form of a break in the spectrum at around the comoving horizon scale at matter-radiation equality

$$r_{\rm H} = 16.0\,[\Omega h^2]^{-1}\ {\rm Mpc} \qquad (1)$$

(as usual, $h \equiv H_0/100\ {\rm km\,s^{-1}\,Mpc^{-1}}$). This number, and all others in the paper, assume $T = 2.726$ K for the CMB temperature (Mather et al. 1994). Since observed wavenumbers come in units of $h\ {\rm Mpc^{-1}}$, the combination $\Omega h$ is measurable, and in practice is estimated by fitting a model of a scale-invariant spectrum modified by the CDM transfer function. According to PD, an approximate 95 per cent confidence range for the apparent value of the density is

$$0.22 < \Omega h|_{\rm apparent} < 0.29 \qquad (2)$$

(allowing for effects of nonlinearity and redshift-space distortions). The problem is that this number appears to be inconsistent with current estimates of $h$ and an $\Omega = 1$ Einstein-de Sitter universe, and a variety of possible solutions have been suggested.

(1) Maybe $\Omega \simeq 0.3$. This is perhaps the most obvious solution observationally, but brings in philosophically worrying fine tunings, either if the universe is open or if vacuum energy provides the remaining fraction of the critical density.

(2) Perhaps $h$ really is around 0.25 (Shanks 1985; Bartlett et al. 1995). This explanation has the merit of simplicity, but would fly in the face of all the experimental evidence.

(3) A primordial spectrum tilted away from scale-invariance alters the fitted apparent density. However, a large degree of tilt is required and this is difficult to reconcile with the CMB studies (Cen et al. 1992; PD).

(4) The fit assumes pure CDM. When baryons are present, what is measured for $\Omega = 1$ is roughly $\Omega h \exp[-2\Omega_{\rm B}]$ (PD). If $h \simeq 0.7$, $\Omega_{\rm B} \gtrsim 0.5$ would be required; it is very hard to see how such a high value can be consistent with primordial nucleosynthesis (e.g. Walker et al. 1991).

(5) One can tinker with the dark-matter content to improve things, and the Mixed Dark Matter model with roughly 20-30% of the dark matter being light neutrinos has received considerable recent attention (e.g. Klypin et al. 1993).

(6) Lastly, the apparent low density can be regarded as in-

forming us that matter-radiation equality must be later than the standard figure. This can be achieved with an enhanced density of relativistic species, probably associated with the decay of a massive particle.

With the exception of (1), all of these are measures designed to save the Einstein-de Sitter universe without abandoning the basic picture of gravitational instability. In this sense, all are tainted with something of an air of desperation. Nevertheless, because of the profound consequences of any disproof of the Einstein-de Sitter model, it is important to explore all avenues thoroughly. The least contrived escape routes seem to be those involving modifications either of the dark matter or of the relativistic content, and it on is this latter possibility that we wish to concentrate. It turns out that the characteristic prediction of such a model is of an enhanced amplitude of fluctuations on small scales, and we use this feature to place limits on the decaying particle involved in the model. In Section 2, we review the model and establish a fitting formula to describe the power spectrum that results. Section 3 assembles the relevant observational constraints in terms of small-scale clustering and the abundances of high-redshift objects.

## 2  CDM WITH A DECAYING PARTICLE

### 2.1  The basic model

Structure formation scenarios incorporating decaying particles have a long history (e.g. Davis et al. 1981; Bardeen, Bond & Efstathiou 1987). However, this model was given a strong boost by experimental work suggesting the possibility of neutrino eigenstates with $m \simeq 17$ keV (e.g. Simpson 1985). Although it is now believed that these results were spurious, the possibility of a mass of this order for e.g. the $\tau$ neutrino is far from being ruled out. Such a neutrino cannot be stable, otherwise it would close the universe many times over. An acceptable present density can be achieved if the massive neutrino decays to products which are relativistic today – either to other massless neutrinos or possibly to some exotic new species. Decays to photons are not allowed for two reasons. First, the COBE results on the lack of CMB spectral distortions severely limit the allowed energy injection prior to recombination (Mather et al. 1994). Second, the relativistic density in photons is observed, and a total relativistic density is conventionally obtained by multiplying by a factor 1.68 to allow for three species of massless neutrinos. The possibility we wish to consider here, however, is that the true relativistic density is higher. If the decays producing this enhanced background occurred at redshifts greater than conventional matter-radiation equality ($z \simeq 24000\,\Omega h^2$), the onset of matter domination would be delayed, and we would have an explanation of the large-scale structure problem. The other constraint on the model is that decay should happen after nucleosynthesis at $z \sim 10^9$, to ensure that the light element abundances are not affected; this leaves a wide range of possible lifetimes and hence masses for the model. We note that Dodelson, Gyuk & Turner (1994) have discussed specifically more complicated nucleosynthesis effects which occur when the decay is contemporaneous with nucleosynthesis (decay lifetime $\sim 10$ seconds), but we will see below that lifetimes this short do not have such interesting consequences for the Mpc-scale fluctuation spectrum.

The analysis presented in this paper follows on from work by Bond & Efstathiou (1991); hereafter BE. They calculated density fluctuations in $\Omega = 1$ models dominated by CDM, in which 17 keV neutrinos having lifetimes between 1 and $10^4$ years decayed to relativistic products. BE derived power spectra which differ from standard CDM in two key ways:

(1) The decay increases the density of relativistic degrees of freedom ('radiation' for short) and so delays the onset of matter-radiation equality. The length scale associated with the Hubble radius at this epoch is correspondingly modified. This modification is parameterized by $\theta$ (Bardeen et al. 1987) which is the ratio of energy density in relativistic species with a decaying particle to that without. Equation (1) now reads

$$r_{\rm H} = 16.0\,[\Omega h^2]^{-1}\,\theta^{1/2}\ {\rm Mpc} \tag{3}$$

and as a result the apparent value of $\Omega h$ (the effective shape parameter for CDM transfer functions) is dependent on the mass and lifetime of the decaying particle. An appropriate choice of these parameters can reconcile the $\Omega h \simeq 0.25$ value which best fits the observed large scale structure with the theoretically favoured $\Omega = 1$ and the observationally implied $h = 0.5 - 0.9$.

(2) If the decay time exceeds $\sim 10 m_{\rm keV}^{-2}$ years the universe can pass through *two* periods of matter domination, the first occurring when the density of undecayed particles exceeds that of relativistic species. This is followed by a phase dominated by the relativistic decay products. The second phase of matter domination arrives when the density of the cold dark matter exceeds that of the decay products. As a result of these processes the power spectrum is characterized by two length scales. Fluctuation growth can occur in the first matter dominated epoch – providing greater small scale power than is generated in the standard CDM scenario, and pushing the formation of sub-cluster size objects to higher redshift.

In this paper we extend the model of BE, allowing both the lifetime and mass of the hypothetical particle to vary. We then constrain these parameters by consideration of structure formation on large and small scales.

### 2.2  Power-spectrum scalings

We now sketch the dependence of $\theta$ on the mass and lifetime of the decaying particle. We adopt the units

$$\begin{aligned} m &\equiv {\rm mass\ /\ keV} \\ \tau &\equiv {\rm decay\ time\ /\ years.} \end{aligned} \tag{4}$$

We will treat the decaying particle as a heavy neutrino, in the sense that the initial number density of the particle will be set equal to that for a massless neutrino. The energy density in decaying particles therefore becomes dominant over the radiation when $mc^2$ is of the order of the radiation energy $kT$. Until this point $\rho \propto T^4$ and the age of the universe is $t \sim (G\rho)^{-1/2} \propto T^{-2} \propto m^{-2}$. Once they are nonrelativistic, the density of heavy neutrinos scales as $\rho \propto a^{-3}$, whereas the density of radiation in the standard massless neutrino model scales as $\rho \propto a^{-4}$. By the time of decay, the massive neutrinos therefore dominate the conventional relativistic density by a factor

$$\frac{\rho_{\rm decay}}{\rho_{\gamma+3\nu}} \simeq \frac{a_{\rm decay}}{a_{\rm eq1}} \simeq \left(\frac{\tau}{t_{\rm eq1}}\right)^{2/3} \propto m^{4/3}\tau^{2/3}. \quad (5)$$

Here $a_{\rm eq1}$ and $a_{\rm decay}$ are the scale factors at the first matter-radiation equality and at decay respectively; $\rho_{\rm decay}$ is the energy density in the relativistic decay products from the massive neutrinos; and $\rho_{\gamma+3\nu}$ is the standard energy density for radiation and three light neutrinos. This assumes an effectively instantaneous decay for the particles at time $\tau$ and a negligible mass for the two other neutrino species. We can now write

$$\theta = \frac{\rho_{\gamma+2\nu} + \rho_{\rm decay}}{\rho_{\gamma+3\nu}} \simeq \frac{1.45}{1.68}\left[1 + x(m^2\tau)^{2/3}\right], \quad (6)$$

with $x$ a dimensionless constant. By a numerical solution of the full equations describing the problem, BE obtained $x \simeq 0.15$.

We now turn to the parameter dependencies of the horizon sizes at the various epochs of matter-radiation equality. The horizon scale at the the time at which the non-relativistic density of the massive neutrinos first becomes dominant is given by equation (3). Now, a massive neutrino has $\Omega h^2 = m/0.095$ in our units (e.g. Kolb & Turner 1990, but scaling to the COBE $T = 2.726$ K); $\theta_{\rm eq1} = 1.45/1.68$; and $\theta_{\rm eq2}$ is as given above. We therefore deduce the sizes of the comoving horizons at the two equality scales:

$$r_{\rm H}^{\rm eq1} = 1.41\, m^{-1}\; {\rm Mpc} \quad (7)$$

$$r_{\rm H}^{\rm eq2} = 14.9\, h^{-2}\left[1 + x(m^2\tau)^{2/3}\right]^{1/2}\; {\rm Mpc} \quad (8)$$

The effect of this is to yield a power spectrum with two characteristic break wavenumbers, of order the reciprocal of the appropriate $r_{\rm H}$. The result can be modeled as the sum of two CDM spectra with differing power-law break lengths and different amplitudes, i.e.

$$\Delta^2(k) = \Delta^2_{\rm LSS}(k) + \alpha^2 \Delta^2_{\rm LSS}(k/\beta) \quad (9)$$

Throughout, we shall express power spectra in dimensionless form:

$$\Delta^2(k) \equiv d\sigma^2/d\ln k = k^3 P(k)/2\pi^2. \quad (10)$$

The expression (9) says that the small-scale power spectrum looks like the CDM model which fits large-scale structure, but with a 'bump' superimposed which is a copy of the large-scale spectrum that has been shifted to smaller scales by a factor $\beta$ and boosted by a factor $\alpha^2$. The large-scale spectrum, $\Delta^2_{\rm LSS}$ would be just the standard BBKS CDM spectrum with an apparent density

$$\Omega h|_{\rm apparent} = \Omega h \left(\frac{1.45}{1.68}\left[1 + x(m^2\tau)^{2/3}\right]\right)^{-1/2}. \quad (11)$$

The shift $\beta$ is just the ratio of the horizon sizes deduced above

$$\beta = 10.6\, m h^{-2}\left[1 + x(m^2\tau)^{2/3}\right]^{1/2}. \quad (12)$$

The 'boost factor' $\alpha$ is more subtle. Until the decay epoch, the spectrum has only the small-scale break, and there can be no growth on these scales during the second period of radiation domination between $a_{\rm decay}$ and $a_{\rm eq2}$. The spectrum on larger scales can grow, which imprints the second break.

If the spectrum has a primordial power-law index $n$, then $\Delta^2 \propto k^{3+n}$ on large scales; the difference in power at the two breaks ($k_1$ & $k_2$, say) is then given by a translational factor $(k_1/k_2)^{3+n}$ and the growth which occurs on large scales during the second period of radiation domination (a factor $[r_{\rm H}^{\rm eq2}/r_{\rm H}^{\rm decay}]^2$ in $\delta\rho/\rho$):

$$\alpha^2 = \frac{[r_{\rm H}^{\rm eq2}/r_{\rm H}^{\rm eq1}]^{3+n}}{[r_{\rm H}^{\rm eq2}/r_{\rm H}^{\rm decay}]^4}. \quad (13)$$

In the scale-invariant $n = 1$ case (assumed hereafter), this is just $\alpha = [r_{\rm H}^{\rm decay}/r_{\rm H}^{\rm eq1}]^2$. To obtain this ratio, note that $r_{\rm H} \propto t^{1/3}$ during the relevant (matter-dominated) phase; since we start at a time $\propto m^{-2}$ and end at $\tau$, we finally obtain

$$\alpha = y\,[m^2\tau]^{2/3}, \quad (14)$$

where $y$ is a further dimensionless constant which must be determined by fitting an exact integration. To obtain the value of this constant, we compared the power spectrum (7) with the BE results for 17 keV neutrinos decaying at 1, 10, $10^2$, $10^3$ and $10^4$ years. These spectra are reproduced in figure 1a. It was found useful to adopt a softening parameter $\gamma$ such that that

$$\Delta^2(k) = \left(\,[\Delta^2_{\rm LSS}(k)]^\gamma + [\alpha^2\Delta^2_{\rm LSS}(k/\beta)]^\gamma\,\right)^{1/\gamma}. \quad (15)$$

This smooths the transition region at which the small scale power of the second term becomes dominant over that of the first. The best fit obtained requires the prefactor for $\alpha$ to be $y = 1.29$, with $\gamma = 0.30$.

To sum up, we have constructed the power spectra for a universe containing $\Omega = 1$ in CDM plus a decaying massive neutrino, including both the correct large-scale shape and the size plus location of the small-scale bump. The requirement for an apparent $\Omega h = 0.25$ fixes $m^2\tau$ if the true Hubble constant is known:

$$m^2\tau \simeq (125 h^2 - 7)^{3/2}, \quad (16)$$

and this also fixes the boost parameter $\alpha$. The only freedom in the model is then the location of the small-scale feature, which depends only on the mass. To illustrate this procedure, figure 1c shows three models with $m^2\tau = 120$, and fixing the apparent $\Omega h$ at 0.25. There is clearly a large range of possibilities for the power at $k \gtrsim 1$, which is what makes this model of interest.

To complete the picture we must allow for the damping of power that occurs due to free streaming of the massive neutrinos whilst they are still relativistic. An analysis of this effect by Bardeen et al. (1986) shows that damping corresponds to approximately Gaussian filtering of the linear power spectrum with $R_f = 2.6(\Omega h^2)^{-1}{\rm Mpc} = 0.247 m^{-1}$ Mpc. In practice, the scales and masses of interest are such that this damping has a negligible effect. Finally, we have seen that there are a number of assumptions which lead to the restriction on $\alpha$ (assumed abundance of decaying particles; scale-invariant spectrum), there will be some advantages to abandoning physical preconceptions on occasion and treating $\alpha$ and $\beta$ as completely free parameters which describe empirically any small-scale features in the power spectrum.

## 3 CONSTRAINTS

### 3.1 Normalization

We now need to place limits on the model by comparing its predictions with observations of structure in the universe. For this, we require normalization for the theoretical power spectra. This can be expressed as the linear theory rms density contrast when averaged over spheres of radius $8h^{-1}$ Mpc, i.e. $\sigma_8$. White, Efstathiou and Frenk (1993) use the observed abundance of rich clusters to deduce a hard allowed range of $\sigma_8 = 0.52 - 0.62$. Direct measurements of $\sigma_8$ from clustering require a knowledge of the bias parameter, and Feldman, Kaiser & Peacock (1994) give $\sigma_8 = 0.91/b - 0.18/b^{1.8}$ from the study of IRAS galaxies. Studies of peculiar velocities and comparison to density fields can yield estimates of $b$; such estimates are summarized in Table 1 of Dekel (1994), and yield $1/b$ from this technique in the range $0.6 - 1.3$, corresponding to $\sigma_8 = 0.47 - 0.89$. The higher values can probably be eliminated by the pairwise random motions of galaxies. This issue was studied thoroughly by Gelb & Bertschinger (1994), who concluded that $\sigma_8 > 0.7$ was untenable, and that even $\sigma_8 = 0.5$ yielded uncomfortably high velocities. We shall adopt $\sigma_8 = 0.6$, which is perhaps at the higher end of the allowed range; since we are looking to see if extra small-scale power is required, it makes sense to be conservative and adopt the highest reasonable normalization for the large-scale mass spectrum. PD give a discussion of the relation between $\Delta^2(k)$ and $\sigma_8$, and argue that $\sigma_8$ largely measures the power at $k \simeq 0.2\,h\mathrm{Mpc}^{-1}$. For BBKS scale-invariant spectra, an accurate numerical fit for the effective wavenumber is

$$k_{\mathrm{eff}}/h\mathrm{Mpc}^{-1} = 0.172 + 0.011[\ln(\Omega h/0.34)]^2. \qquad (17)$$

Finally, therefore, we adopt $\Delta^2(0.18\,h\ \mathrm{Mpc}^{-1}) = 0.6^2$ as our normalization.

To evolve these linear power spectra to the present day ($z = 0$) involves non-linearities which can alter the power spectrum significantly at small scales. In evolving the $\Delta^2(k) \gtrsim 1$ portion of the power spectrum we use the formulae of PD (derived from the work of Hamilton et al. 1991, which was based on N-body simulations on the relevant scales) to incorporate the effects of non-linear evolution. Whilst this correction increases the small-scale power for shallow power-law regions it actually removes power in regions with a strong $k$-dependence. Figures 1b & 1d show the non-linear correction applied to the spectra of 1a & 1c respectively.

### 3.2 High redshift objects

One way of constraining the small-scale ($k \gtrsim 0.2h\ \mathrm{Mpc}^{-1}$) power is to require that it is sufficient to form the observed abundances of high redshift objects, in particular quasars, radio galaxies and damped Lyman $\alpha$ systems. Radio galaxies have a well-defined mass, but this is rather large; they are also a rare population, so that current data do not set a very strong constraint on fluctuation spectra (Peacock 1994). Quasars are a more numerous population, but the ease with which observed quasar abundances can be attained is very much dependent on the (uncertain) mass assumed (e.g. Efstathiou & Rees 1988; Haenhelt 1993) and we have avoided their use here.

More stringent constraints can be derived from recent deep measurements of damped Ly$\alpha$ systems with HI column densities greater than $\sim 2 \times 10^{20}$ cm$^{-2}$ (Lanzetta et al. 1991) These have been used by several authors to investigate rival dark matter models. If the fraction of baryons in the virialized dark matter halos equals the global value $\Omega_B$, then these data can be used to infer the total fraction of matter that has collapsed into bound structures at high redshifts (Ma & Bertschinger 1994, Mo & Miralda-Escudé 1994). The highest measurement at $\langle z \rangle \simeq 3.2$ implies $\Omega_{\mathrm{HI}} \simeq 0.005$, and hence a collapsed fraction of $\simeq 10\%$ if $\Omega_B = 0.05$. Here we apply the Walker et al. (1991) constraint to the baryon density, namely $\Omega_B h^2 = 0.0125 \pm 0.0025$.

The assumption here will be that the damped Ly$\alpha$ systems are the progenitors of present day spiral galaxies. Evidence for this view has been provided by absorption in lensed quasar systems, showing the systems to be of galactic dimensions (Wolfe et al. 1992). Furthermore the baryon mass inferred in present day galaxies is comparable to that of the damped Ly$\alpha$ systems at $z \simeq 3$. The photoionizing background prevents systems with circular velocities of less than about 50 km s$^{-1}$ cooling sufficiently to form bound systems (e.g. Efstathiou 1992). We follow Mo & Miralda-Escudé (1994) and use this conservative velocity limit to estimate the minimum mass of object that the Ly$\alpha$ measurements detect. Virial equilibrium for a halo of mass $M$ and radius $r_v$ demands

$$v_c^2 = \frac{GM}{r_v} \qquad (18)$$

For a spherically collapsed object this velocity can be converted directly into a Lagrangian comoving radius containing this mass (White et al. 1993)

$$r_0 = \frac{2^{1/2} v_c}{H_0 \Omega^{1/2}(1+z)^{1/2}(1+178\Omega^{-0.6})^{1/6}} \qquad (19)$$

The values introduced above require $r_0 > 0.15\,h^{-1}$ Mpc.

To use the measurement to constrain our candidate power spectra we employ the well-known formalism of Press & Schechter (1974) which gives a collapsed fraction $\Omega_c$ above some mass scale $M_{\mathrm{coll}}$, given a redshift $z$ and a means of computing the $z = 0$ rms density contrast as a function of mass $\sigma(M, z = 0)$:

$$\Omega_c(> M_{\mathrm{coll}}, z) = 1 - \mathrm{erf}\left[\delta_c(1+z)/\sqrt{2}\sigma(M_{\mathrm{coll}}, z=0)\right]. \qquad (20)$$

$\sigma(M)$ can be evaluated by filtering the power spectrum on the required scale. Here we use a spherical 'top hat' filter of radius $R_T$ (for which $f(k) = 3(\sin y - y\cos y)/y^3$ with $y = kR_T$) with a corresponding mass of $4\pi\rho R_T^3/3$. Our lower mass limit corresponding to $r_0 = 0.15 h^{-1}$ Mpc is therefore $M_{\mathrm{coll}} = 10^{9.6} h^{-1} \mathrm{M}_\odot$. $\delta_c$ is the critical overdensity required for collapse, which for a 'top-hat' overdensity undergoing spherical collapse is 1.686. This canonical value has recently received support from N-body simulations on relevant scales by Ma & Bertschinger (1994).

Our power spectra are therefore required to produce an $\Omega_c$ which *exceeds* the Ly $\alpha$ collapsed fraction at all $z$, for $M > 10^{9.6} h^{-1} \mathrm{M}_\odot$. The true collapsed fraction could be higher than the observed one because some large fraction of the baryonic material in collapsed haloes could be ionized or in stars. Figure 2 shows how various models fare in meeting this demand. The points with errors are the HI fraction in damped Lyα systems from Lanzetta et al. (1993) and Storrie-Lombardi et al. (1995). Standard $\Omega h = 0.5$ CDM, curve (c), successfully attains a sufficient collapsed fraction at the high redshift end. Models that agree with the APM at large scales do less well. (b) $\Omega h = 0.25$ CDM and (a) MDM with $\Omega_\nu = 0.25$ both fall short of having enough small scale power. The $\Omega h = 0.5$ CDM + relativistic decay model (d) for $\tau = 10$ yr and $m = 3.5$ keV has ample small scale power thanks to the extra bump from the first epoch of matter domination.

How seriously should these constraints be taken? A pure $\Omega h = 0.25$ spectrum does not fail to fit the data by a very large amount, and it is not implausible that such a model could be rescued by tweaking the assumptions in the calculations – perhaps most readily by assuming a larger $\Omega_B$. Furthermore, the data themselves may not be definitive. The inferred column density of a damped Lyα system is exponentially sensitive to its velocity width, and this can easily be artificially enhanced by superposition of Lyα forest systems in the wings. Storrie-Lombardi et al. (1995) also suggest that the $\langle z \rangle = 3.2$ Lanzetta point may be slightly too high. We therefore cannot claim that the need for extra small-scale power is rigorously established. Nevertheless, because it is unlikely that all collapsed HI can escape ionization, it is valuable to explore models which allow a significant increase of the collapsed fraction at high $z$.

### 3.3 Galaxy clustering

The above constraints require only some minimum level of power; however, we are not at liberty to exceed this minimum by too large a factor. One limitation is provided by modelling the infrared Tully-Fisher relation between galaxy luminosity and circular velocity in the scenario of interest. A number of authors (e.g. Cole et al 1994) have shown that standard $\Omega h = 0.5$, $\sigma_8 \simeq 0.7$ CDM has too much power on small-scales, yielding, for a given luminosity, circular velocities which are 60% larger than is observed. Without a relatively complicated modelling of galaxy formation we cannot subject our candidate spectra to the same test.

A more straightforward upper limit to the power can be provided by comparison of the candidate power spectra with the observed small-scale power spectrum, best determined by angular deprojection of the APM galaxy survey (Baugh & Efstathiou 1993), who give data down to $k \simeq 8h$ Mpc$^{-1}$. The degree of small-scale bias relating the power spectrum of mass to that of the observed light is not known with any great accuracy, but the mass-to-light ratios of clusters strongly encourage us to believe that the small-scale clustering of light must exceed that of mass, if $\Omega = 1$. Although some models have been advocated in which this would not be true (e.g. the paper by Couchman & Carlberg 1992 on $b = 1$ standard CDM), it is reasonable to regard such a situation as observationally unacceptable. We will therefore set a conservative upper limit to the allowed degree of small-scale power by demanding that the theoretical nonlinear power spectrum of the mass at no point exceeds that of the light, as measured by the deprojection of the APM angular clustering. As with the Tully-Fisher relation $\Omega h = 0.5$ CDM with our chosen normalization fails this test.

## 4 ALLOWED MODELS

### 4.1 Limits on parameters

The limits on the decaying neutrino models derived by the above methods are summarized in figures 3 & 4.

Figures 3(i) & 3(ii) are more 'empirical' representations of the results in that they show the values of $\alpha$ and $\beta$ which give a permissible power spectrum in light of the Lyman-$\alpha$ and APM constraints. 3(i) & 3(ii) differ in the assumed value of the Hubble constant, $h = 0.5$ and $0.75$ respectively. The baryon fraction depends on $h$ and is correspondingly modified. The large scale portion of the theoretical power is fixed at $\Omega h = 0.25$ CDM and the small-scale high $k$ portion alone is varied. A band of acceptable values results, which narrows as $h$ is increased, through the change in $\Omega_B$. As $\beta$ and $\alpha$ are dependent on the combination $(m^2\tau)^{2/3}$ selecting a particular value of $h$ effectively fixes the 'growth' parameter $\alpha$. The dotted lines on the figure indicate the growth allowed when $m^2\tau$ is such that $\Omega h = 0.5$ and $\Omega h = 0.75$ are matched on to the large-scale observations. Within the usual range discussed for $h$ therefore the model is constrained to have a shift parameter $\beta \simeq 80 - 2500$ and a growth factor $\alpha \simeq 30 - 80$.

Figures 4(i) & 4(ii) translate the results of figures 3(i) & 3(ii) from general form, to specifically apply to the decaying neutrino model, identifying regions in the mass-lifetime plane ruled out by the constraints discussed above. Figure 4(i) refers to an $\Omega h = 0.50$ model with $\Omega_B = 0.050$, figure 4(ii) to $\Omega h = 0.75$, $\Omega_B = 0.022$. The region disallowed by the Lyα structure formation requirement at z=3.2 is indicated by the darker shaded areas. In order not to exceed the small-scale APM curve, the parameters must lie somewhere in the plane away from the lighter shaded region. The dashed line from top-left to bottom-right of each plot represents $m^2\tau \simeq 120$ and $500$, the values required to reconcile $\Omega h_{\text{apparent}} = 0.25$ with $\Omega h_{\text{true}} = 0.50$ and $0.75$. We can summarize the conclusions from these figures as follows:

$$\begin{aligned}\Omega h = 0.50 &\Rightarrow 0.5 < m < 30 \text{ keV}, \ 0.2 < \tau < 500 \text{ yr} \\ \Omega h = 0.75 &\Rightarrow 2.0 < m < 30 \text{ keV}, \ 0.5 < \tau < 100 \text{ yr}.\end{aligned} \quad (21)$$

Are these parameter values physically plausible? The dot-dashed line in figures 4(i) & 4(ii) shows the equation

$$m^5\tau = 3 \times 10^4 \text{ keV}^5\text{yrs}, \quad (22)$$

which is the form of a naive prediction for the relation between mass and lifetime for a particle which decays via the weak interaction. This takes the usual $E^5$ scaling of weak-interaction cross sections and scales to the decay of the $\mu$ lepton. In neither case does this 'muon-decay' line cross the large-scale structure line in a permitted region; if the model is to be considered plausible, the decay physics involved must be more exotic.

### 4.2 Early reionization

The existence of significant small scale power in decaying particle CDM can allow structures of mass $10^5 - 10^8 h^{-1} M_\odot$ to form earlier than in the standard model and so may permit early reionization of the intergalactic medium. We have performed a brief analysis based on the method of Tegmark, Silk and Blanchard (1994). They estimate a parameter $f_{\text{net}}$, the net efficiency of ionization processes from stars, and calculate $f_s(M)$ the collapsed fraction of the universe (that gives rise to the star formation). These they relate to an ionization fraction $\chi$ such that

$$\chi \simeq 3.8 \times 10^5 f_{\rm net} f_s \qquad (23)$$

Tegmark et al. give a range of $f_{\rm net}$ values they believe permissible, dubbing the top of the range 'optimistic' (in the sense of promoting early reionization), the bottom of the range 'pessimistic' and the median value 'middle-of-the-road'. Once $f_{net}$ is set in this manner the requirement of 100% reionization ($\chi > 1$) becomes a condition on $f_s$ which can be calculated by the same Press-Schechter method as was used to analyze the Lyman-$\alpha$ constraint. In order to get a feel for the sort of redshifts at which reionization could occur in our model we set $m^2\tau \simeq 500$, the value needed to give an apparent $\Omega h = 0.25$ in the case where $\Omega = 1$ and $h = 0.75$. We then vary the mass within the range 0.01 to 100 keV, $\tau$ being set at $500/m^2$. $f_s$ is calculated for the resulting power spectra and used to derive the redshift at which reionization is complete.

In figure 5 we show the expected reionization redshifts for each of the $f_{\rm net}$ values suggested. The mass scale for the Press-Schechter integration (cf. eqn. 20) relates to the masses of the first galaxies to form and there is a considerable range of possibilities from $10^5 h^{-1} M_\odot$ (Couchman and Rees 1986) to $10^7 - 10^8 h^{-1} M_\odot$ (Blanchard et al. 1992). Fortunately the power spectra are relatively flat in this range and the collapsed fraction is relatively insensitive to $M_{\rm coll}$. Our results relate to the mass scale $10^7 h^{-1} M_\odot$. In rough agreement with the Tegmark et al. results, typical reionization redshifts are in the range $\sim 10 - 200$ for an optimistic or middle-of-the-road $f_{\rm net}$. A pessimistic $f_{\rm net}$ permits early reionization only around $m \sim 1$ keV. CMB photons will be significantly scattered by the reionized plasma if the optical depth between $z_{\rm ion}$ and $z = 0$ is $\simeq 1$. To obtain this requires $z_{\rm ion} \simeq 50$ (e.g. Padmanabhan 1993) which can occur in the relativistic decay model for $m \sim 2 - 10$ keV – within our allowed range. Such a result may well be of importance in wiping out details of the last scattering surface on angular scales around $\sim 1°$ (e.g. White, Scott & Silk 1994). We note here that White, Gelmini and Silk (1994) predict a sensitive dependence of degree scale CMB anisotropies on the parameters of the decaying particle model. They suggest that $m^2\tau$ may, in principle, be determined from the CMB, with a large 1° bump for models with large $m^2\tau$. Such a feature is excluded observationally, but we have shown that the small scale power in this model may well cause sufficiently early reionization to significantly complicate the measurements in this range.

## 5  SUMMARY

We have demonstrated that the parameters of the $\Omega = 1$ decaying particle + CDM model can be constrained by small-scale power-spectrum requirements. Whilst we must generate sufficient power at $\sim 0.1 h^{-1}$ Mpc scales to significantly affect the observed number of high-$z$ damped Ly$\alpha$ systems, the existence of APM clustering data on comparable scales limits how much power can be added. Successful models of this general kind are therefore quite tightly constrained. Although present data cannot be claimed to provide definitive proof for a small-scale feature in the power spectrum, interesting consequences are predicted for future data on high-redshift objects.

Within an acceptable range for $h$, the mass of any decaying neutrino is constrained to lie between 0.5 to 30 keV and the lifetime to lie between 0.2 and 500 years. Structure formation can commence sufficiently early in this scenario to permit early reionization of the IGM. For a range of allowed parameters the reionization will inevitably occur sufficiently early to modify CMB fluctuations on scales of $\sim 1°$ and below.

Although it thus has some attractive features, the model is not without its difficulties. Since we retain the Einstein-de Sitter universe, the problem of short Hubble times is not evaded, and may ultimately prove fatal. Furthermore, as with all high-density models, a convincing mechanism for biased galaxy formation needs to be supplied. However, the merit of the picture studied here is that it is reasonably well constrained and so testable directly in terms of particle physics. If the Einstein-de Sitter model is to be saved, this is arguably the least contrived way to do it.

## 6  ACKNOWLEDGEMENTS

SJM acknowledges the support of a SERC/PPARC research studentship. We thank Dick Hunstead and Rocky Kolb for helpful discussions, and particularly Michael Turner for saving us from error at an early stage of this investigation.

## FIGURE CAPTIONS

**Figure 1**
(a) Best fits to the Bond & Efstathiou (1991) power spectra obtained via the scalings of section 2.2. The mass is set at 17 keV and the lifetime varied from 0 to $10^4$ years. The 0 yrs curve is equivalent to standard CDM (BBKS 1986). There is progressively more large scale power as the lifetime is increased. h=0.5 throughout.
(b) The power spectra of (a) subjected to the non-linear correction of Peacock & Dodds (1994; PD).
(c) Three power spectra with parameters satisfying $m^2\tau = 120$ and hence fulfilling the large scale structure requirement (eqn. 2) for $\Omega h_{\text{true}} = 0.5$. A range of small scale behaviour is still possible.
(d) The power spectra of (b) subjected to the non-linear correction of PD. The points with error bars are the power spectrum obtained by angular deprojection of the APM galaxy survey (Baugh & Efstathiou 1993).

**Figure 2** Collapsed baryon fractions (in terms of critical density) for four dark matter models. (a) Mixed dark matter with $\Omega_{\text{HDM}}$=25 %, and h = 0.5 (b) CDM with $\Omega h = 0.25$, (c) CDM with $\Omega h = 0.5$, (d) CDM + relativistic decay model with $\Omega h = 0.5$, $\tau = 10$ years and $m = 3.5$ keV. The points with $1\sigma$ error bars are damped Ly$\alpha$ system densities adopted from Lanzetta (1993) apart from the $z = 4$ point, taken from Storrie-Lombardi et al. (1995). Only models (c) and (d) have enough small scale power to account for the collapsed fraction at $z \simeq 3.2$, though model (c) is ruled out by large scale observations. We have assumed $\Omega_{\text{B}} = 0.05$ throughout, each model being normalized to $\sigma_8 = 0.6$ at $z = 0$. All these comparisons assume $h = 0.5$; since $\Omega_{\text{HI}}$ scales as $h^{-1}$, a higher true Hubble constant would lower the data points and make it easier for some models to satisfy the constraints.

**Figure 3** Constraints on the $\alpha$ and $\beta$ parameters introduced in section 2.2. For a given value of $\beta$, $\alpha$ must lie *above* the Ly$\alpha$ line for a sufficient collapsed fraction to form at $z = 3.2$. $\alpha$ must however lie *below* the APM curve to avoid exceeding the APM power spectrum for $k > 1h$ Mpc$^{-1}$. A band of permitted $\alpha$ and $\beta$ values results (unshaded region). Furthermore if $m^2\tau$ is fixed to match the large scale structure then $\alpha = $ const. This is shown as the dashed dotted-line in each case:-
(i) $\Omega_{\text{B}} = 0.050$, $\Omega h = 0.50$
(ii) $\Omega_{\text{B}} = 0.022$, $\Omega h = 0.75$

**Figure 4** Constraints on mass and lifetime of hypothetical decaying neutrino. Limits are based on the damped Ly$\alpha$ fraction at redshift $z = 3.2$ (darker shaded region forbidden). Values which would exceed the APM power spectrum for $k > 1h$ Mpc$^{-1}$ occupy the lighter shaded region.
(i) Model constraints for $\Omega h = 0.5$, $\Omega_{\text{B}} = 0.050$. The dashed line shows the large-scale structure constraint for $\Omega h = 0.5$, the dot-dashed line the $m^5\tau \simeq const$ law for decaying heavy leptons.
(ii) Model constraints for $\Omega h = 0.75$, $\Omega_{\text{B}} = 0.022$. The dashed line shows the large-scale structure constraint for $\Omega h = 0.75$, the dot-dashed line the $m^5\tau \simeq const$ law for decaying heavy leptons.

**Figure 5** Reionization redshifts as a function of decaying particle mass for the case $\Omega h = 0.75$, i.e. $m^2\tau \simeq 500$. We take the mass of the earliest galaxies to form to be $10^7 h^{-1} \text{M}_\odot$ though the results are similar for the range $10^5 - 10^8 h^{-1} \text{M}_\odot$. The labels 'optimistic', 'middle-of-the-road' and 'pessimistic' (abbreviated to opt., m.o.r., and pess.) refer to the efficiency of reionization by stars and are discussed in the text. Reionization occurs in the range $z \simeq 10 - 200$ for most of the parameter space investigated.